\documentclass[a4paper, 11pt]{article}
\usepackage{}
\usepackage{amsfonts}
\usepackage{dsfont}
\usepackage{mathrsfs}
\usepackage{amssymb}
\usepackage{bbm}
\usepackage{epsfig}
\usepackage{amsmath}
\usepackage{appendix}
\usepackage[english]{babel}

\def\Reals{\mathop{\hbox{\mit I\kern-.2em R}}\nolimits}

\def\Complexes{{\hbox{\mit C\kern-.46em
            \vrule depth 0ex height 1.4ex width .05em\kern.41em}}}

\newtheorem{them}{Theorem}[section]
\newtheorem{defn}{Definition}[section]

\newtheorem{lem}{Lemma}[section]
\newtheorem{pro}{Proposition}[section]
\newtheorem{remark}{Remark}[section]
\newtheorem{rem}{Remark}[section]

\title{\bf  An Approximate Projected Consensus Algorithm \\ for Computing Intersection of Convex Sets\footnote{This work is partially
supported by the NNSF of China under Grant 61174071,  the Knut and Alice Wallenberg Foundation and the Swedish Research Council.}}
\date{}

\author{Youcheng Lou\thanks{Y. Lou and Y. Hong are with Key Laboratory of Systems and Control, Institute of Systems
       Science, Chinese Academy of Sciences, Beijing 100190, China.
       Email: {\tt\small  louyoucheng@amss.ac.cn, yghong@iss.ac.cn}}, Guodong Shi\thanks{G. Shi and K. H. Johansson are with ACCESS Linnaeus Centre, School of Electrical Engineering,
Royal Institute of Technology, Stockholm 10044, Sweden.
       Email: {\tt\small guodongs@kth.se, kallej@ee.kth.se}}, Karl Henrik Johansson and Yiguang Hong}

\begin{document}

\maketitle

\begin{abstract}
In this paper, we  propose an approximate projected consensus algorithm  for a network to cooperatively compute the intersection of convex sets.   Instead of assuming the exact convex projection proposed in the literature, we allow each node to compute an approximate projection and communicate it to its neighbors. The communication graph is directed and time-varying. Nodes update their states by weighted averaging.  Projection accuracy conditions are presented for the considered algorithm. They indicate how much projection accuracy is required to ensure global consensus to a point in the intersection set when the communication graph is uniformly jointly strongly connected. We show that $\pi/4$ is a critical angle error of the projection approximation to ensure a bounded state. A numerical example  indicates that this approximate projected consensus algorithm may  achieve better performance than the exact projected consensus algorithm in some cases.
\end{abstract}


\section{Introduction}
In recent years, dynamics and control on large-scale networks have drawn increasing research attention in different  areas including engineering, computer
science, and social science. Cooperative
control of a group of autonomous agents fully employs local information exchange and distributed protocol design to accomplish collective tasks such as consensus, formation, and aggregation \cite{tsi,jad03,caoming1,ren,mar, Wang,tantac,sabertac,shi09,shi11}. Moreover, in parallel computation, load-balance problems require realtime balance of the load from different  computing resources \cite{cs2,cs3}. Additionally, a central problem of opinion dynamics in social networks is how the agreement is achieved via individual belief exchange processes \cite{degroot,daron}. A fundamental question in these problems is,  how consensus can be guaranteed based on local information exchange, time-varying node interconnection and limited knowledge of the global objective.

Various distributed optimization problems arise for consensus with particular optimization purpose in practice. Minimizing a sum of  convex functions,
where each component is known only to a particular node, has
attracted much attention, due to its simple formulation and wide applications \cite{
bj08,rabbat, ram07, nedic1, nedic2,nedic4, shi1, shi2, lu1, lu2, moura,joh09,ram}.
The key idea is that properly designed distributed control protocols
or computation algorithms can lead to a collective optimization,
based on simple exchanged information and individual optimum
observation.  Subgradient-based incremental methods were established
via deterministic  or randomized iteration, where each node is
assumed to be able to compute a local subgradient value of its
objective function \cite{rabbat, ram07, nedic1, bj08,joh09,ram}.
Non-subgradient-based methods also showed up in the literature. For
instance, a non-gradient-based algorithm was proposed, where each
node starts at its own optimal solution and updates using a pairwise
equalizing protocol \cite{lu1,lu2}, and later an augmented
Lagrangian method was introduced in \cite{moura}.

In particular, if the optimal solution set of its own objective can
be obtained for each node, the considered optimization problem is
then converted to a set intersection computation problem when we
additionally assume there is a nonempty intersection among all
solution sets \cite{shi1, shi2, nedic2}. In fact, convex intersection computation problem is a
classical problem in the optimization study
\cite{inter1,inter2,inter3}. The so-called  ``alternating projection algorithm" was a standard centralized solution, where projection  is carried out alternatively  onto each set \cite{inter1, inter2, inter3}. Then the ``projected consensus algorithm"  was presented   as a decentralized version of alternating projection algorithm, where each node alternatively projects onto its own set and averages with its neighbors, and comprehensive convergence analysis was given for this projected algorithm under time-varying directed interconnections in \cite{nedic2}.  Following this work,  a flip-coin algorithm was
introduced when each node randomly chooses projection or averaging by
Bernoulli processes, and almost sure convergence was shown for the
system to reach an optimal consensus in \cite{shi1}. A dynamical system
solution was given in \cite{shi2}, where the network reaches a global
optimal consensus by a simple continuous-time control. In all these algorithms, each node needs to know the exact convex projection of its current state onto its objective set \cite{shi1,shi2,nedic2}.

However, in practice, the exact convex projection is usually hard to compute due to the common environmental noise and computation inaccuracy.  In this paper, we therefore propose an approximate projected consensus algorithm  to solve the convex intersection computation problem. Instead of assuming the exact convex projection, we allow each node to just compute an approximate projection point which locates in the intersection of the convex cone generated by the current state and all directions with the exact projection direction less than some angle and the half-space containing the current state with its boundary being a supporting hyperplane to its own set at its exact projection point onto its set.  The communication graph is supposed to be directed and time-varying. With uniformly jointly strongly connected conditions, we show that the whole network can achieve a global consensus within the intersection of all convex sets when sufficient projection accuracy can be guaranteed. For a special approximate projection case when the nodes can get the exact direction of the projection, a necessary and sufficient condition is given on how much projection accuracy is critical to ensure a global intersection computation. A numerical example is also given, and surprisingly, the approximate projected consensus algorithm  sometimes achieves better performance for convergence than the exact projected consensus algorithm.

The paper is organized as follows. Section 2 gives some basic
concepts on graph theory and convex analysis. Section 3 introduces the network model and formulates the problem of interest. Section 4 presents the main results and convergence analysis for the considered algorithm. Section 5 focuses on the discussion on the critical angle error of the approximate projection. Section 6 gives a numerical example and finally, Section 7 shows some concluding remarks.

\section{Preliminaries }

In this section, we introduce preliminary knowledge on graph theory \cite{God} and convex analysis \cite{Roc}.


A directed graph (digraph) ${\mathcal{G}}=(\mathcal{V},\mathcal{E}, A)$ consists of node
set $\mathcal{V}=\{1,2,...,n\}$, arc set
$\mathcal{E}\subseteq \mathcal{V} \times \mathcal{V}$ and an
 adjacency matrix $A=[a_{ij}]_{n\times n}$
with nonnegative adjacency elements $a_{ij}$.
The element $a_{ij}$
of matrix $A$ associated with arc $(i,j)$ is positive if and only if $(i,j) \in \mathcal{E}$.
$\mathcal{N}_i$ denotes the set of neighbors of node $i$, that is,
$\mathcal{N}_i=\{j\in\mathcal{V}|(i,j)\in\mathcal{E}\}$.
In this paper, we assume $(i,i)\in\mathcal{E}$ for all $i$.
A path from $i$ to $j$ in digraph $\mathcal{G}$
is a sequence
$(i_0, i_1),(i_1, i_2),..., (i_{p-1}, i_p)$ of arcs
with $i_0=i$ and $i_p=j.$ $\mathcal{G}$ is said to be strongly
connected if there exists a path from $i$ to $j$ for each
pair of nodes $i,j\in\mathcal{V}$.


A function $f(\cdot): R^m\rightarrow R$ is said to be convex if $
f(\lambda x + (1-\lambda)y)\leq\lambda f(x)
+ (1-\lambda)f(y)$ for any $x, y \in R^m$ and $ 0<\lambda<1.$
A function $f$ is said to be concave if $-f$ is convex.
A set $K\subseteq R^m$ is said to be convex if $\lambda x + (1-\lambda)y\in K$ for any $x, y \in K$ and $ 0< \lambda <1$
and is said to be a convex cone if $\lambda_1 x + \lambda_2y\in K$ for any $x, y \in K$ and $\lambda_1,\lambda_2\geq 0$.
 For a
closed convex set $K$ in $R^m$, we can associate to any $x\in R^m$ a
unique element $P_K(x)\in K$ satisfying $|x-P_K(x)|= \inf_{y\in
K}|x-y|$, which is denoted as $|x|_K,$ where $|\cdot|$
denotes the Euclidean norm and $P_K$ is the projection operator onto
$K$.

For a
closed convex set $K$, if $x\not\in K$, then by the supporting
hyperplane theorem, there is a supporting hyperplane to $K$ at $P_{K}(x)$.
The angle between vectors $a$ and $b$ is denoted as ${\rm Ang}(a,b)\in[0,\pi]$ for which
$\cos {\rm Ang}(a,b)=\langle a, b\rangle/(|a||b|)$, where $\langle a, b\rangle$ denotes the Euclidean inner product of vectors $a$ and $b$.

We cite a lemma for the following analysis (see Example 3.16 in \cite{Boyd}, pp. 88).

\begin{lem}\label{lem1}
$f(z)=|z|_K$ is a convex function, where $K$ is a closed convex set in $R^m$.
\end{lem}

The following properties hold for the projection operator $P_K$.
Here (i) is the standard non-expansiveness property for convex projection; (ii) comes from exercise 1.2 (c) in \cite{Cla} (pp. 23)
and (iii) is a special case of proposition 1.3 in \cite{Cla} (pp. 24).
\begin{lem}\label{lem2}
Let $K$ be a closed convex set in $R^m$. Then
\begin{align}
&(i)\quad |P_K(x)-P_K(y)|\leq|x-y|\;\mbox{for any}\; x\; \mbox{and}\; y;\nonumber\\
&(ii)\quad \big||x|_K-|y|_K\big|\leq|x-y|\;\mbox{ for any}\; x \;\mbox{and}\; y;\nonumber\\
&(iii)\quad P_K(\lambda x+(1-\lambda) P_K(x))=P_K(x)\;\mbox{for any}\; x \;\mbox{and}\; 0<\lambda<1.\nonumber
\end{align}
\end{lem}

The next lemma can be found in \cite{shi1}.
\begin{lem}\label{lem3}
Let $K$ and $K_0\subseteq K$ be two closed convex sets.
We have
$$|P_K(x)|^2_{K_0}+|x|^2_K\leq |x|^2_{K_0}\;\mbox{ for any}\; x.$$
\end{lem}

\section{Problem Formulation}

In this section, we introduce the intersection computation problem and the
approximate projected consensus algorithm.

Consider a network consisting of $n$ agents with node set $\mathcal{V}=\{1,2,...,n\}$. Each node $i$ is associated with a set $X_i\subseteq R^m$ and
set $X_i$ is known only by node $i$. The intersection of all these sets is nonempty, i.e., $\bigcap^n_{i=1}X_i\neq\emptyset$.
Let us denote $X_0=\bigcap^n_{i=1}X_i$. The target of the network is to find a point in $X_0$ in a distributed way.
For $X_i,i=1,...,n$, we use the following
assumption:

\noindent {\bf A1} {\it (Convexity)} $X_i,i=1,...,n$, are closed convex sets.

\begin{rem}
The intersection computation problem can be equivalently converted into the following distributed optimization problem:
the
objective of this group of $n$ agents is not only to achieve a
consensus, but also to cooperatively solve
$$
\min_{x\in R^m}\sum^n_{i=1}f_i(x),
$$
where $f_i: R^m\rightarrow R$ is the convex cost function of agent $i$ and can be known only by agent $i$.
Here $X_i=\{y|f_i(y)=\min_{x\in R^m}f_i(x)\},1\leq i\leq n$ are closed convex sets, which are assumed to be
nonempty and have a nonempty intersection.
\end{rem}

\subsection{Communication Graphs}
The communication over the network is modeled as a sequence of directed graphs,
$\mathcal{G}_k=(\mathcal{V},\mathcal{E}(k),A(k)),k\geq 0$. We say node $j$ is a neighbor of node $i$
at time $k$ if there is an arc $(i,j)\in\mathcal{E}(k)$, where $a_{ij}(k)$ represents its weight.
Let $\mathcal{N}_i(k)$ denote the set of neighbors of agent $i$ at time $k$.
We introduce an assumption on the weights \cite{nedic1,shi1}.

\noindent {\bf A2} {\it (Weights Rule)} (i) $\sum_{j\in\mathcal{N}_i(k)}a_{ij}(k)=1$ for all $i$ and $k$.

(ii) There exists a constant $0<\eta<1$ such that $a_{ij}(k)\geq \eta$ for
all $i,k$ and $j\in\mathcal{N}_i(k)$.

For the connectivity of the communication graphs, we introduce the following assumption \cite{shi2,nedic2}.

\noindent {\bf A3} {\it (Connectivity)}
 The communication graph is uniformly
jointly strongly connected $(UJSC)$, i.e., there exists a positive
integer $T$ such that $\mathcal{G}([k,k+T))$ is strongly connected
for $k\geq 0$, where $\mathcal{G}([k,k+T))$ denotes the union
graph with node set $\mathcal{V}$ and arc set $\bigcup_{k\leq s<
k+T}\mathcal{E}(s)$.

\subsection{Approximate Projection}
Projection methods have been widely used to solve various problems,
including projected consensus \cite{nedic2}, the convex
intersection computation \cite{inter2, inter3} and distributed computation \cite{Ber1}. In the most literature,
the projection point $P_{K}(z)$ of $z$ onto closed convex set $K$ is
required to achieve desired convergence, but in practice it is hard to be obtained and often is
computed approximately. Here is the definition of
approximate projection.
\begin{defn}
\label{app}
Suppose $K\subseteq R^m$ is a closed convex set and $0<\theta<\pi/2$.
Define
\begin{align}
&\mathbf{C}_K(v,\theta)=v+\big\{ z|\;\langle z, P_K(v)-v\rangle\geq |z||v|_K\cos\theta \big\};\nonumber\\
&\mathbf{H}^+_K(v)=\big\{z|\;\langle v-P_K(v), z\rangle\geq\langle v-P_K(v), P_K(v)\rangle\big\}.\nonumber
\end{align}
as a convex cone and a half-space, respectively.
The approximate projection $\mathscr{P}^a_K(v,\theta)$
of point $v$ onto $K$ with approximate angle $\theta$ is defined as the following set:
\begin{equation}
\label{ang}\mathscr{P}^a_K(v, \theta)=
\left\{
\begin{array}{ll}
\mathbf{C}_K(v,\theta)\bigcap \mathbf{H}^+_K(v),\;\mbox{if}\;\;v\not\in K\\
\{v\},\quad\quad\quad\quad\quad\quad\;\; \mbox{if}\;\;v\in K\\
\end{array}
\right.
\end{equation}
\end{defn}

In fact,
$\mathbf{C}_K(v,\theta)-v$ is a convex cone generated by
all vectors having angle with $P_K(v)-v$ less than $\theta$
and $\mathbf{H}^+_K(v)$ is the half-space containing point $v$ with
$$\mathbf{H}_K(v):=\big\{z|\;\langle v-P_K(v), z\rangle=\langle v-P_K(v), P_K(v)\rangle\big\}$$
being a supporting hyperplane to $K$ at $P_K(v)$.

\begin{figure}[!htbp]
\centering
\includegraphics[width=3.4in]{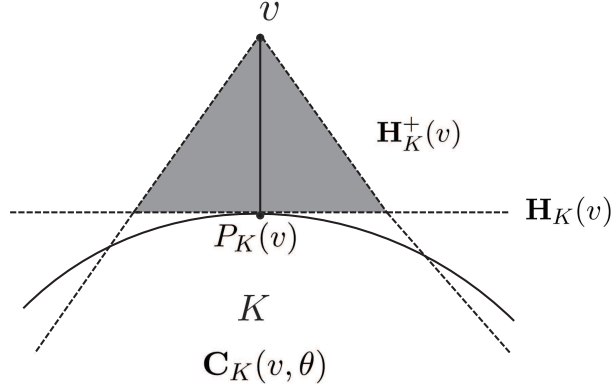}
\caption{The set marked by the shaded area is the approximate projection of $v$ onto $K$.}
\end{figure}

\begin{defn}\label{def2}
The supporting approximate projection $\mathscr{P}^{sa}_K(v, \theta)$ of point $v$ onto $K$ with approximate angle $\theta$ is defined as
$$
\mathscr{P}^{sa}_K(v, \theta)=
\begin{cases}
 \mathbf{C}_K(v,\theta)\bigcap \mathbf{H}_K(v),\;\mbox{if}\;\;v\not\in K\\
 \{v\},\quad\quad\quad\quad\quad\quad\;\; \mbox{if}\;\;v\in K
 \end{cases}
$$
\end{defn}

\begin{remark}
Exact projection may be difficult to obtain in practice, due to possible data quantization and  limitation of the estimate. This is why we introduce approximate projection, as a relaxed estimate to the real projection value.
\end{remark}
According to Definition \ref{def2}, for any $y\in\mathscr{P}^a_K(v, \theta)$, we can associate $y$ with
 $\hat y\in\mathscr{P}^{sa}_K(v, \theta)$ such that
\begin{align}\label{a}
y= (1-\beta) v+\beta\hat y\;\;\mbox{for some}\;\;0\leq \beta\leq 1.
\end{align}
Moreover, it is easy to see that if $y\neq v$, $\hat y$ satisfying (\ref{a}) is unique.

\subsection{Distributed Iterative Algorithm}

To solve the intersection computation problem, we propose the following
approximate projected consensus algorithm:
\begin{equation}
\label{6}
 x_i(k+1)=\sum_{j\in\mathcal{N}_i(k)}a_{ij}(k)P^a_j(k),\;i=1,...,n,
 \end{equation}
where $P^a_i(k)\in \mathscr{P}^a_{X_i}(x_i(k), \theta_k)$ for all $i$ and $k$.

According to the definition of supporting approximate projection, there exist $0\leq\alpha_{i,k}\leq 1$
and $P^{sa}_i(k)\in\mathscr{P}^{sa}_{X_i}(x_i(k), \theta_k)$ such that
\begin{align}\label{7}
P^a_i(k)=(1-\alpha_{i,k})x_i(k)+\alpha_{i,k}P^{sa}_i(k).
\end{align}

Combining with (\ref{6}) and (\ref{7}), we have
\begin{align}\label{8}
x_i(k+1)=\sum_{j\in\mathcal{N}_i(k)}a_{ij}(k)\Big((1-\alpha_{j,k})x_j(k)+\alpha_{j,k}P^{sa}_j(k)\Big),
\end{align}
where if $x_i(k)\not\in X_i$,
$P^{sa}_i(k)\in\mathbf{H}_{X_i}(x_i(k))$ and ${\rm Ang}(P^{sa}_i(k)-x_i(k),P_{X_i}(x_i(k))-x_i(k))\leq\theta_k$.

\begin{figure}[!htbp]
\centering
\includegraphics[width=3.6in]{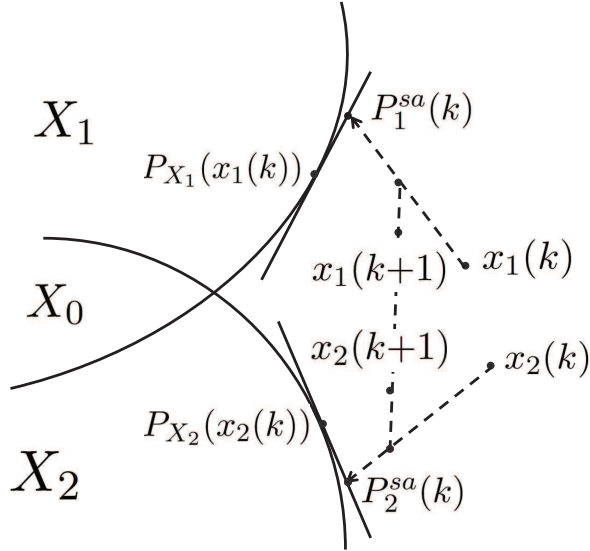}
\caption{The approximate projected consensus algorithm.}
\end{figure}

\begin{remark}
Note that the ``projected consensus
algorithm" presented  in \cite{nedic2} is a special case with $\alpha_{i,k}\equiv1$ and $\theta_k\equiv0$ of the approximate projected consensus
algorithm discussed in this paper.
\end{remark}

We illustrate the iteration process of the algorithm in Figure 2.
For the approximate angle $\theta_k$, we use the following assumption.

\noindent {\bf A4}  $0\leq\theta_k\leq \theta^*<\pi/2$ for all $k$.

In this paper, we are interested in whether an optimal consensus can be achieved or not, as defined  in the following definition.

\begin{defn}
\label{def1} A global optimal consensus is achieved for the approximate projected consensus
algorithm if, for any initial condition $x(0)\in R^{nm}$,
there exists $x^*\in X_0$ such that
$$
\lim_{k\rightarrow\infty}x_i(k)=x^*,\;i=1,...,n.\;
$$
\end{defn}

Note that a global optimal consensus $x^*$ necessarily belongs to the intersection set $X_0$. Hence, a possible algorithm
to find a point in $X_0$ is to employ the approximate projected consensus algorithm (\ref{6}). We next discuss the convergence
of this algorithm.

\section{Main Results and Convergence Analysis}
In this section, we present the main results and convergence analysis.

Denote $\alpha^-_k=\min_{1\leq i\leq n}\alpha_{i,k}$ and $\alpha^+_k=\max_{1\leq i\leq n}\alpha_{i,k}, k\geq0.$
\begin{them}
\label{thm1} Suppose {\bf A1}--{\bf A4} hold.
Global optimal consensus is achieved for the approximate projected consensus
algorithm
if $\sum^{\infty}_{k=0}\alpha^-_k=\infty$ and $\sum^{\infty}_{k=0}\alpha^+_k\theta_k<\infty$.
\end{them}

To investigate the necessity
of the divergence condition in Theorem \ref{thm1}, we impose another assumption on the boundedness of the $n$ sets $X_i,i=1,...,n$.

\noindent {\bf A5} {\it (Bounded Sets)} $X_i,i=1,...,n$, are bounded sets.

Then the following conclusion holds showing a necessary projection accuracy condition.
\begin{them}
\label{thm2} Suppose {\bf A1}--{\bf A5} hold, $\theta_k\equiv0$ and $\alpha^+_k<1$ for all $k$.
Global optimal consensus is achieved for the approximate projected consensus
algorithm only if $\sum^{\infty}_{k=0}\alpha^+_k=\infty$.
In fact, if $\sum^{\infty}_{k=0}\alpha^+_k<\infty$, then for initial condition $x_i(0)=z^*, i=1,...,n$ with
$|z^*|_{X_0}>\sum^\infty_{k=0}\alpha^+_kd^*/\prod^\infty_{k=0}(1-\alpha^+_k)$, there is
$y^*=y^*(z^*)\not\in X_0$
such that $\lim_{k\rightarrow\infty}x_i(k)=y^*$ for all $i$, where
$d^*=\sup_{\omega_1,\omega_2\in\bigcup^n_{i=1}X_i}|\omega_1-\omega_2|$.
\end{them}

\begin{rem}
Suppose {\bf A1}--{\bf A5} hold, $\theta_k\equiv0$ and
there exists a sequence $\{\alpha_k\}^\infty_{k=0}$ with $\alpha_k<1$ for all $k$ such that $\alpha_{i,k}=\alpha_{j,k}=\alpha_k$ for all
$i,j$ and $k$. Then from Theorems 1 and 2, we have that global optimal consensus is achieved for the approximate projected consensus
algorithm if
and only if $\sum^{\infty}_{k=0}\alpha_k=\infty$.
\end{rem}

\begin{rem}
Compared to the convergence results given in \cite{nedic2}, Theorems \ref{thm1} and \ref{thm2} do not require the doubly stochastic assumption
on the weights $a_{ij}(k)$ ($\sum^n_{j=1}a_{ij}(k)=\sum^n_{j=1}a_{ji}(k)=1$ for all $i,k$). This is important because double stochasticity is hard to guarantee for the arc weights in a distributed way, especially when the communication is directed.

Moreover, the connectivity assumption in \cite{nedic2} requires that $\mathcal
{G}([k,k+T))$ is a fixed graph for sufficiently large $k$, which is more restrictive than our UJSC assumption. However, the assumption in \cite{nedic2} can be relaxed to UJSC graphs, as indicated in a comment by the authors.
\end{rem}

\subsection{Lemmas}

We establish several useful lemmas in this subsection.

Let $\{x_i(k)\}^{\infty}_{k=0}$ be the states of node $i$ generated by (\ref{6}), $i=1,...,n$.
Denote $|x(k)|_{X_0}=(|x_1(k)|_{X_0}...|x_n(k)|_{X_0})^T$, $y(k)=(y_1(k)...y_n(k))^T$ with $y_i(k)=|x_i(k)|_{X_0}-\sqrt{|x_i(k)|^2_{X_0}-|x_i(k)|^2_{X_i}}$.
Denoting $D_k={\rm diag}\{\alpha_{1,k},\alpha_{2,k},...,\alpha_{n,k}\}$,
the following lemma holds.

\begin{lem} Suppose \noindent {\bf A1} holds.
For all $k\geq s$, we have
\begin{align}\label{5}
 |x(k+1)|_{X_0}\leq A(k)|x(k)|_{X_0}-A(k)D_ky(k)+\tan\theta_kA(k)D_k|x(k)|_{X_0}.
\end{align}
\end{lem}
Proof.
According to Lemma \ref{lem1}, (\ref{8}) implies
\begin{align}\label{ineqa}
|x_i&(k+1)|_{X_0}\leq\sum_{j\in\mathcal{N}_i(k)}a_{ij}(k)\Big((1-\alpha_{i,k})|x_j(k)|_{X_0}+\alpha_{i,k}|P^{sa}_j(k)|_{X_0}\Big).
\end{align}

By Lemma \ref{lem2} (ii), we have
\begin{align}\label{435}
|P^{sa}_j(k)|_{X_0}&\leq\big|P^{sa}_j(k)-P_{X_j}(x_j(k))\big|+|P_{X_j}(x_j(k))|_{X_0}.
\end{align}
The definition of $P^{sa}_j(k)$ ensures that
\begin{align}\label{pro}
\big|P^{sa}_j(k)-P_{X_j}(x_j(k))\big|\leq\tan\theta_k|x_j(k)|_{X_j}.
\end{align}
Moreover, it follows from Lemma \ref{lem3} that for any $j\in\mathcal{V}$,
\begin{align}\label{017}
|P_{X_j}(x_j(k))|_{X_0}&\leq\sqrt{|x_j(k)|^2_{X_0}-|x_j(k)|^2_{X_j}}.
\end{align}

It follows from (\ref{ineqa}), (\ref{435}), (\ref{pro}), (\ref{017}) and the relation $|x_j(k)|_{X_j}\leq|x_j(k)|_{X_0}$ that
\begin{align}
|x_i(k+1)|_{X_0}\leq\sum_{j\in\mathcal{N}_i(k)}&a_{ij}(k)\Big((1-\alpha_{j,k})|x_j(k)|_{X_0}+\alpha_{j,k}
\sqrt{|x_j(k)|^2_{X_0}-|x_j(k)|^2_{X_j}}\Big)\nonumber\\
\label{178}
&+\tan\theta_k\sum_{j\in\mathcal{N}_i(k)}a_{ij}(k)\alpha_{j,k}|x_j(k)|_{X_0}\\
=\sum_{j\in\mathcal{N}_i(k)}&a_{ij}(k)\Big(|x_j(k)|_{X_0}-\alpha_{j,k}\big(|x_j(k)|_{X_0}-
\sqrt{|x_j(k)|^2_{X_0}-|x_j(k)|^2_{X_j}}\big)\Big)\nonumber\\
\label{177}
&+\tan\theta_k\sum_{j\in\mathcal{N}_i(k)}a_{ij}(k)\alpha_{j,k}|x_j(k)|_{X_0}.
\end{align}
Then the conclusion follows.
\hfill$\square$

It is easy to find that
$\tan\theta\leq (\tan\theta^*/\theta^*)\theta$ for $0\leq\theta\leq \theta^*$.
Thus, if $\sum^{\infty}_{k=0}\alpha^+_k\theta_k<\infty$, then $\sum^{\infty}_{k=0}\alpha^+_k\tan\theta_k<\infty$.

\begin{lem}
\label{bound} Suppose \noindent {\bf A1} and \noindent {\bf A4} hold.
If $\sum^{\infty}_{k=0}\alpha^+_k\theta_k<\infty$, $\{x_i(k)\}^{\infty}_{k=0}$ is bounded for all $i$.
\end{lem}
Proof. Lemma 1 (b) in \cite{nedic2} states that for closed convex $K$ and any $x\in R^m$,
\begin{align}\label{13}
|P_{K}(x)-y|^2\leq |x-y|^2-|x|^2_{K}\;\mbox{for all}\; y\in K.
\end{align}
By taking $K=X_j$ and $z\in X_0\subseteq X_j$, (\ref{13}) leads to
\begin{align}\label{202}
|P_{X_j}(x_j(k))-z|\leq\sqrt{|x_j(k)-z|^2-|x_j(k)|^2_{X_j}}.
\end{align}

By considering $|x_i(k+1)-z|$ instead of $|x_i(k+1)|_{X_0}$, following similar procedures with (\ref{ineqa}), (\ref{435}), (\ref{pro}),
and substituting (\ref{017}) with (\ref{202}), we can show that for any $i\in \mathcal{V}$,
\begin{align}\label{201}
|x_i(k+1)-z|&\leq\sum_{j\in\mathcal{N}_i(k)}a_{ij}(k)|x_j(k)-z|+\tan\theta_k\sum_{j\in\mathcal{N}_i(k)}a_{ij}(k)\alpha_{j,k}|x_j(k)-z|
\nonumber\\
&\quad\quad-\sum_{j\in\mathcal{N}_i(k)}a_{ij}(k)\alpha_{j,k}\Big(|x_j(k)-z|-\sqrt{|x_j(k)-z|^2-|x_j(k)|^2_{X_j}}\Big).
\end{align}
By dropping the non positive term on the righthand in (\ref{201}), we have
\begin{align}\label{188}
\max_{1\leq i\leq n}|x_i(k+1)-z|\leq(1+\alpha^+_k\tan\theta_k)\max_{1\leq i\leq n}|x_i(k)-z|.
\end{align}

Therefore, for all $k$, we have
\begin{align}\label{437}
\max_{1\leq i\leq n}|x_i(k+1)-z|&\leq\prod^k_{l=0}\big(1+\alpha^+_l\tan\theta_l\big)\max_{1\leq i\leq n}|x_i(0)-z|\nonumber\\
&\leq e^{\sum^{k}_{l=0}\alpha^+_l\tan\theta_l}\max_{1\leq i\leq n}|x_i(0)-z|\nonumber\\
&\leq e^{\sum^{\infty}_{l=0}\alpha^+_l\tan\theta_l}\max_{1\leq i\leq n}|x_i(0)-z|,
\end{align}
where the second inequality follows from $1+z\leq e^z$ for $z\geq 0$. Then the conclusion follows.
\hfill$\square$

The next lemma is a special case
of various random versions, for example, see Lemma 11 in \cite{Plo} (pp. 50).
Here we give a simple proof.
\begin{lem}
\label{4.1} Let $\{a_k\}^{\infty}_{k=0}$ and $\{b_k\}^{\infty}_{k=0}$ be non-negative
 sequences with $\sum^{\infty}_{k=0}b_k<\infty$. Suppose
$$
a_{k+1}\leq a_k+b_k\mbox{ for all}\; k
$$
Then $\lim_{k\rightarrow\infty}a_k$ is a finite number.
\end{lem}
Proof. Define a new sequence $\{c_k\}^{\infty}_{k=1}$ with $c_k=a_k-\sum^{k-1}_{l=0}b_l$, which
is bounded since $\{a_k\}^{\infty}_{k=1}$ and $\{\sum^{k-1}_{l=0}b_l\}^{\infty}_{k=1}$ are bounded.
Moreover, $c_{k+1}\leq c_k$ for $k\geq 1$ implies that
$\lim_{k\rightarrow\infty}c_k$ is a finite number. The conclusion follows from
$\lim_{k\rightarrow\infty}a_k=\lim_{k\rightarrow\infty}c_k+\sum^{\infty}_{l=0}b_l$.
\hfill$\square$

\begin{lem}
\label{lemA1} Suppose \noindent {\bf A1} and \noindent {\bf A4} hold.
If $\sum^{\infty}_{k=0}\alpha^+_k\theta_k<\infty$, the following limit exists
$$
\vartheta:=\lim_{k\rightarrow\infty}\max_{1\leq i\leq n}|x_i(k)|_{X_0}.
$$
\end{lem}
Proof.
Take $z\in X_0$. Based on (\ref{177}) and (\ref{437}), we have
\begin{align}
\max_{1\leq i\leq n}|x_i(k+1)|_{X_0}\leq\max_{1\leq i\leq n}|x_i(k)|_{X_0}+\alpha^+_k\tan\theta_ke^{\sum^{\infty}_{l=0}\alpha^+_l\tan\theta_l}\max_{1\leq i\leq n}|x_i(0)-z|.\nonumber
\end{align}
The conclusion follows from the last inequality and Lemma \ref{4.1}.
\hfill$\square$

Denote
$$
\eta^+_i=\limsup_{k\rightarrow\infty}|x_i(k)|_{X_0},\;\;\eta^-_i=\liminf_{k\to
\infty}|x_i(k)|_{X_0},\;i\in\mathcal{V}.
$$
Obviously, $0\leq \eta^-_i\leq \eta^+_i\leq \vartheta$ for all $i$.
\begin{lem}
\label{lemA2} Suppose \noindent {\bf A1}--\noindent {\bf A4} hold. If $\sum^{\infty}_{k=0}\alpha^+_k\theta_k<\infty$ and there
exists some agent $i_0\in\mathcal{V}$ such
that
$\eta^-_{i_0}<\vartheta$, then $\vartheta=0$.
\end{lem}
\noindent {Proof:}
Motivated by the idea of Lemma 4.3 in \cite{shi2}, we prove this lemma by contradiction.

Denote $$\varrho_i=(\eta^-_{i}+\eta^+_{i})/2,\; i\in\mathcal{V}.$$
Since $\eta^-_{i_0}<\vartheta$,
there exists an increasing sequence $\{k_l\}^{\infty}_{l=0}$ such that
$|x_{i_0}(k_l)|_{X_0}\leq\varrho_{i_0}<\vartheta$
for $l\geq 0$. Moreover, for any $\varepsilon>0$, there exist
$K_0=K_0(\varepsilon)$ such that
$|x_i(k)|_{X_0}\leq \vartheta+\varepsilon$ and $d_0\sum^{\infty}_{k=K_0}\alpha^+_k\theta_k\leq \varepsilon$ for $k\geq K_0$ and all $i$,
where
\begin{align}\label{18}
d_0=(\tan\theta^*/\theta^*)\sup_{1\leq i\leq n,k\geq 0}|x_i(k)|_{X_0},
\end{align}
 which is finite by Lemma \ref{bound}.
Without loss of generality, we assume $k_0\geq K_0$.

Based on inequality (\ref{177}), we have
\begin{align}
|x_{i_0}(k_0+1)|_{X_0}\leq\sum_{j\in\mathcal{N}_{i_0}(k_0)\backslash i_0}&a_{i_0j}(k_0)|x_j(k_0)|_{X_0}+a_{i_0i_0}(k_0)|x_{i_0}(k_0)|_{X_0}+d_0\alpha^+_{k_0}\theta_{k_0}.\nonumber
\end{align}
Therefore,
$|x_{i_0}(k_0+1)|_{X_0}\leq (1-\eta)(\vartheta+\varepsilon)+\eta\varrho_{i_0}+d_0\alpha^+_{k_0}\theta_{k_0}$
and then
\begin{align}
|x_{i_0}(k_0+2)|_{X_0}&\leq (1-\eta)(\vartheta+\varepsilon)+\eta[(1-\eta)(\vartheta+\varepsilon)+d_0\alpha^+_{k_0}\theta_{k_0}]+\eta\varrho_{i_0}+d_0\alpha^+_{k_0+1}\theta_{k_0+1}
\nonumber\\
&\leq(1-\eta^2)(\vartheta+\varepsilon)+\eta^2\varrho_{i_0}+d_0\sum^{k_0+1}_{k=k_0}\alpha^+_k\theta_k.\nonumber
\end{align}
Similarly, we can show by induction that for $r\geq 1$,
\begin{align}
|x_{i_0}(k_0+r)|_{X_0}\leq(1-\eta^r)(\vartheta+\varepsilon)+\eta^r\varrho_{i_0}+d_0\sum^{k_0+r-1}_{k=k_0}\alpha^+_k\theta_k.\nonumber
\end{align}

Since the communication graph is UJSC, there exist agent $i_1\neq i_0$
and time $k^1_0\in[k_0,k_0+T)$ such that $(i_1,i_0)\in\mathcal{E}(k^1_0)$.
As the above estimate for $|x_{i_0}(k_0+r)|_{X_0}$ with $x_{i_0}(k_0)$,
by considering $|x_{i_1}(k^1_0+r)|_{X_0}$ with $|x_{i_0}(k^1_0)|_{X_0}$, we can show similarly that for $r\geq 1$,
\begin{align}
|x_{i_1}(k^1_0+r)|_{X_0}&\leq(1-\eta^r)(\vartheta+\varepsilon)+\eta^r|x_{i_0}(k^1_0)|_{X_0}+d_0\sum^{k^1_0+r-1}_{k=k^1_0}\alpha^+_k\theta_k\nonumber\\
&\leq(1-\eta^{k^1_0-k_0+r})(\vartheta+\varepsilon)+\eta^{k^1_0-k_0+r}\varrho_{i_0}+d_0\sum^{k^1_0+r-1}_{k=k_0}\alpha^+_k\theta_k.\nonumber
\end{align}

Repeating the previous procedure on intervals $[k_0+pT,k_0+(p+1)T),1\leq p\leq n-2$, we can obtain nodes $\{i_2,i_3,...,i_{n-1}\}$
such that $\{i_j,0\leq j\leq n-1\}=\mathcal{V}$ and
\begin{align}
\max_{1\leq i\leq n}|x_{i}(k_0+\hat T)|_{X_0}&\leq(1-\eta^{\hat T})
(\vartheta+\varepsilon)+\eta^{\hat T}\varrho_{i_0}+d_0\sum^{\infty}_{k=k_0}\alpha^+_k\theta_k\nonumber\\
&\leq(1-\eta^{\hat T})
(\vartheta+\varepsilon)+\eta^{\hat T}\varrho_{i_0}+\varepsilon,\nonumber
\end{align}
where $\hat T=(n-1)T$.
Moreover, we can make similar analysis for $k_1,k_2,...$
and obtain that for $l\geq 0$,
$$
\max_{1\leq i\leq n}|x_{i}(k_l+\hat T)|_{X_0}\leq(1-\eta^{\hat T})
(\vartheta+\varepsilon)+\eta^{\hat T}\varrho_{i_0}+\varepsilon,
$$
which yields a contradiction since
$(1-\eta^{\hat T})(\vartheta+\varepsilon)+\eta^{\hat T}\varrho_{i_0}+\varepsilon<\vartheta$
provided that $\varepsilon$ is sufficiently small.
\hfill$\square$

We introduce  transition matrices
$$\Phi(k,s)=A(k)\cdots A(s+1)A(s)\;\;\mbox{for all}\; k\; \mbox{and} \;s\; \mbox{with}\; k\geq s.$$
Recall that $\eta$ and $T$ are defined in \noindent {\bf A2} and \noindent {\bf A3}, respectively and
$\hat T=(n-1)T$.
The next lemma generalizes Lemma 2 in \cite{nedic1} on the lower bound of the entries of the transition matrices.

\begin{lem}
\label{lemA3}
Suppose \noindent {\bf A2} and \noindent {\bf A3} hold. Then
$\Phi(k,s)_{ij}\geq\eta^{\hat T}$ for all $i,j,s$ and $k\geq s+\hat T-1$.
\end{lem}
\noindent {Proof:}
By Lemma 2 in \cite{nedic1}, $\Phi(s+\hat T-1,s)_{ij}\geq \eta^{\hat T}$ for all $i,j$ and $s\geq 0$.
Moreover, according to \noindent {\bf A2} (i), $\sum^n_{l=1}A(k)_{il}=1$ and then $\sum^n_{l=1}\Phi(k,s+\hat T)_{il}=1$ for all $i,k$ and $s$.
Thus, for all $i,j$ and $k\geq s+\hat T-1$,
\begin{align}
\Phi(k,s)_{ij}&=\Big(\Phi\big(k,s+\hat T\big)\Phi\big(s+\hat T-1,s\big)\Big)_{ij}\nonumber\\
&\geq\sum^n_{l=1}\Phi\big(k,s+\hat T\big)_{il}\min_{1\leq p,q\leq n}\Phi\big(s+\hat T-1,s\big)_{pq}\nonumber\\
&=\eta^{\hat T}.\nonumber
\end{align}
The conclusion follows.
\hfill$\square$

\begin{lem}
\label{lemA4}
$$\frac{1}{n}\sum^n_{i=1}\sqrt{ v^2_0-v^2_i}\leq \sqrt{v^2_0-\Big(\frac{\sum^n_{i=1}v_i}{n}\Big)^2},$$
where $v_0\geq v_i\geq0$ for all $i$.
\end{lem}
\noindent {Proof:}
The conclusion follows from that
$f(z)=\sqrt{c^2-z^2}$ with domain $[-c, c]$ is a concave function,
where $c>0$.
\hfill$\square$

Consider the following consensus model with noise $w_i$,
\begin{equation}
\label{12}z_i(k+1)=\sum_{j\in\mathcal{N}_i(k)}b_{ij}(k)z_j(k)+w_i(k),\;i=1,...,n,
\end{equation}
where the weights $b_{ij}(k),i,j\in\mathcal{V},k\geq 0$ satisfy {\bf A2}.
The consensus is said to be achieved for system (\ref{12}) if for any initial conditions,
$\lim_{k\rightarrow\infty}|z_i(k)-z_j(k)|=0$ for all $1\leq i,j\leq n$.
The next lemma can be obtained from Theorem 1 in \cite{Wang}.

\begin{lem}
\label{lem5} If the communication graph of system (\ref{12}) is
UJSC with $\lim_{k\rightarrow\infty}w_i(k)=0$ for all $i$, then
the consensus is achieved for system (\ref{12}).
\end{lem}

\subsection{Proofs}
In this subsection, we present the proofs of Theorems \ref{thm1} and \ref{thm2}.

\subsubsection{Proof of Theorem \ref{thm1}}
 Rewrite (\ref{8}) as
\begin{align}\label{consen}
x_i(k+1)&=\sum_{j\in\mathcal{N}_i(k)}a_{ij}(k)x_j(k)+\Big(\sum_{j\in\mathcal{N}_i(k)}a_{ij}(k)\alpha_{j,k}\big(P_{X_j}(x_j(k))-x_j(k)\big)\Big)\nonumber\\
&\quad+\sum_{j\in\mathcal{N}_i(k)}a_{ij}(k)\alpha_{j,k}\Big(P^{sa}_j(k)-P_{X_j}(x_j(k))\Big).
\end{align}
Based on (\ref{pro}), the second and the third term in (\ref{consen}) are not greater than
\begin{align}\label{21}
\max_{1\leq i\leq n}\alpha_{i,k}|x_i(k)|_{X_i}+\alpha^+_k\tan\theta_k\max_{1\leq i\leq n}|x_i(k)|_{X_i}.
\end{align}

Note that
$\vartheta=0$ leads to
$\lim_{k\rightarrow\infty}\max_{1\leq i\leq n}|x_i(k)|_{X_i}\leq
\lim_{k\rightarrow\infty}\max_{1\leq i\leq n}|x_i(k)|_{X_0}=0$ and then the term in (\ref{21}) tends to zero as $k\rightarrow\infty$.
 Therefore, by applying Lemma \ref{lem5} for (\ref{consen}),
we have that if $\vartheta=0$, then the consensus is achieved.

Moreover, we claim that if $\vartheta=0$ and the consensus is achieved, then all agents will converge to a point in
$X_0$. Since $\{x_i(k)\}^\infty_{k=0},i=1,...,n$ are bounded by Lemma \ref{bound} and the consensus is achieved, there is $x^*\in X_0$ and a subsequence $\{k_l\}^{\infty}_{l=1}$
such that $\lim_{l\rightarrow\infty}x_i(k_l)=x^*$ for all $i$.
By similar analysis with (\ref{437}), we have
\begin{align}
\max_{1\leq i\leq n}|x_i(k)-x^*|\leq e^{\sum^{\infty}_{p=k_l}\alpha^+_p\tan\theta_p}\max_{1\leq i\leq n}|x_i(k_l)-x^*|
\leq e^{\sum^{\infty}_{p=0}\alpha^+_p\tan\theta_p}\max_{1\leq i\leq n}|x_i(k_l)-x^*|\nonumber
\end{align}
for $k\geq k_l$, which implies $\lim_{k\rightarrow\infty}x_i(k)=x^*$ for all $i$.

If there exists some agent $i_0$ such
that $\eta^-_{i_0}<\vartheta$, then  by Lemma \ref{lemA2}, $\vartheta=0$.
Therefore, we only need to prove
\begin{align}
\vartheta=0\mbox{ when }\eta^+_{i}=\eta^-_{i}=\vartheta\mbox{  for all }i, \nonumber
\end{align}
which shall be proven by contradiction. If $\vartheta>0$,
then for any $\varepsilon>0$, there exist $K_1=K_1(\varepsilon)$
such that $|x_i(k)|_{X_0}\leq \vartheta+\varepsilon$ and $d_0\alpha^+_k\theta_k\leq \varepsilon$ for $k\geq
K_1$ and all $i$. We complete the proof by the following two steps.

\begin{itemize}
\item [(i)] Suppose $\eta^+_{i}=\eta^-_{i}=\vartheta$ for all $i$. The consensus is achieved: $\lim_{k\rightarrow\infty}|x_i(k)-x_j(k)|=0$ for all $i,j$.

Denote
$$
\varsigma_i=\limsup_{k\rightarrow\infty}\alpha_{i,k}|x_i(k)|_{X_i},\;i\in\mathcal{V}.
$$
We prove $\varsigma_i=0$ for all $i$ by contradiction. If
there exists some agent $i_0$ such that $\varsigma_{i_0}>0$, then there is an
increasing time subsequence $\{k_l\}^{\infty}_{l=1}$ with $k_1\geq K_1$ such that
$\alpha_{i_0, k_l}|x_{i_0}(k_l)|_{X_{i_0}}\geq c\varsigma_{i_0}$ for all $l$ and some
$0<c<1$. Therefore, by (\ref{178})
\begin{align}\label{17}
|x_{i_0}(k_l+1)|_{X_0}&\leq a_{i_0i_0}(k)
\Big((1-\alpha_{i_0, k_l})|x_{i_0}(k_l)|_{X_0}+
\alpha_{i_0, k_l}\sqrt{|x_{i_0}(k_l)|^2_{X_0}-|x_{i_0}(k_l)|^2_{X_{i_0}}}\Big)\nonumber\\
&\quad\quad\quad\quad+\sum_{j\in\mathcal{N}_{i_0}(k_l)\backslash i_0}a_{{i_0}j}(k_l)|x_j(k_l)|_{X_0}+d_0\alpha^+_{k_l}\theta_{k_l}\nonumber\\
&\leq\eta\Big((1-\alpha_{i_0, k_l})(\vartheta+\varepsilon)
+\sqrt{\alpha^2_{i_0, k_l}(\vartheta+\varepsilon)^2-c^2\varsigma^2_{i_0}}\Big)+(1-\eta)(\vartheta+\varepsilon)+\varepsilon\nonumber\\
&=(1-\eta\alpha_{i_0, k_l})(\vartheta+\varepsilon)+\eta
\sqrt{\alpha^2_{i_0, k_l}(\vartheta+\varepsilon)^2-c^2\varsigma^2_{i_0}}+\varepsilon,
\end{align}
where $d_0$ is the one in (\ref{18}), which yields a contradiction since the right hand side of (\ref{17})
is less than $\vartheta$
for sufficiently small $\varepsilon$ and sufficiently large $l$.

Thus,
$\lim_{k\rightarrow\infty}\alpha_{i, k}|x_{i}(k)|_{X_i}=0$ for all $i$. Moreover, since $\sum^{\infty}_{k=0}\alpha^+_k\theta_k<\infty$, $\lim_{k\rightarrow\infty}\alpha^+_k\tan\theta_k$ $\leq(\tan\theta^*/\theta^*)\lim_{k\rightarrow\infty}\alpha^+_k\theta_k=0$.
The two preceding conclusions and the boundedness of $\{x_i(k)\}^{\infty}_{k=0}$ imply that the term in (\ref{21}) tends to zero and then the consensus is achieved by applying Lemma \ref{lem5} for (\ref{consen}) again.

\item [(ii)] Suppose $\eta^+_{i}=\eta^-_{i}=\vartheta$ for all $i$. All agents converge to the nonempty intersection set: $
\lim_{k\rightarrow\infty}|x_i(k)|_{X_0}=0$ for all $i.$

Denote
$$\delta=\liminf_{k\rightarrow\infty}\sum^n_{i=1}|x_{i}(k)|_{X_{i}}.$$
We prove that $\delta=0$ by contradiction. Otherwise, suppose $\delta>0$.

By (\ref{5}), we obtain for $k\geq s$,
\begin{align}\label{16}
 |x(k+1)|_{X_0}&\leq \Phi(k,s)|x(s)|_{X_0}-\sum^k_{l=s}\Phi(k,l)D_ly(l)+d_0\sum^k_{l=s}\alpha^+_l\theta_l\nonumber\\
 &=\Phi(k,s)|x(s)|_{X_0}-\sum^{k-\hat T+1}_{l=s}\Phi(k,l)D_ly(l)\nonumber\\
 &\quad\quad\quad\quad\quad\quad\quad\quad-\sum^{k}_{l=k-\hat T+2}\Phi(k,l)D_ly(l)+d_0\sum^k_{l=s}\alpha^+_{l}\theta_l,
\end{align}
where $\hat T=(n-1)T$. By dropping the third term (nonpositive) on the right-hand side in (\ref{16}), we obtain
\begin{align}\label{19}
|x(k+1)|_{X_0}\leq&\Phi(k,s)|x(s)|_{X_0}-\sum^{k-\hat T+1}_{l=s}\Phi(k,l)D_ly(l)+d_0\sum^k_{l=s}\alpha^+_{l}\theta_l.
\end{align}

For $\bar\varepsilon=\delta^2/(4n^2\vartheta+2\delta)$, there exists sufficiently large $K_2$ such that $\sum^n_{i=1}|x_{i}(k)|_{X_{i}}>\delta-\bar\varepsilon$ and $\vartheta-\bar\varepsilon\leq|x_i(k)|_{X_0}\leq \vartheta+\bar\varepsilon$ for $k\geq K_2$.
For $k\geq K_2$, we have
\begin{align}
\sum^n_{i=1}\sqrt{|x_i(k)|^2_{X_0}-|x_i(k)|^2_{X_i}}&\leq \sum^n_{i=1}\sqrt{(\vartheta+\bar\varepsilon)^2-|x_i(k)|^2_{X_i}}\nonumber\\
&\leq n\sqrt{(\vartheta+\bar\varepsilon)^2-\Big(\big(\sum^n_{i=1}|x_i(k)|_{X_i}\big)/n\Big)^2}\nonumber\\
&\leq n\sqrt{(\vartheta+\bar\varepsilon)^2-\big((\delta-\bar\varepsilon)/n\big)^2},\nonumber
\end{align}
where the second inequality follows from Lemma \ref{lemA4} and then
\begin{align}
\sum^n_{i=1}\Big(|x_i(k)|_{X_0}-\sqrt{|x_i(k)|^2_{X_0}-|x_i(k)|^2_{X_i}}\Big)
&\geq n\Big(\vartheta-\bar\varepsilon-\sqrt{(\vartheta+\bar\varepsilon)^2-\big((\delta-\bar\varepsilon)/n\big)^2}\Big)\nonumber\\
&:=\zeta>0.\nonumber
\end{align}
Namely, $\sum^n_{i=1}y_i(l)\geq\zeta$ for $l\geq K_2$.
Combining the preceding inequality with Lemma \ref{lemA3}, we have that every component
of $\Phi(k,l)D_ly(l)$ is not less than $\eta^{\hat T}\zeta\alpha^-_l$ for all $K_2\leq l\leq k-\hat T+1$ and all $k\geq K_2+\hat T-1$.
Then by (\ref{19}) with taking $s=K_2$, we obtain
\begin{align}\label{20}
|x(k+1)|_{X_0}\leq\Phi(k,K_2)|x(K_2)|_{X_0}-\eta^{\hat
T}\zeta\sum^{k-\hat T+1}_{l=K_2}\alpha^-_l\textbf{1}+d_0\sum^k_{l=K_2}\alpha^+_l\theta_l,
\end{align}
where $\textbf{1}$ is the vector of all ones.
Observing that $\sum^{\infty}_{l=K_2}\alpha^-_l=\infty$, $\sum^{\infty}_{l=K_2}\alpha^+_l\theta_l<\infty$,  and
noticing $\lim_{k\rightarrow\infty}|x(k)|_{X_0}=\vartheta\textbf{1}$,
a contradiction arises  by taking the limit as $k\rightarrow\infty$ in (\ref{20}).

Therefore,
$\delta=0$,
that is, there is a subsequence $\{k_l\}^{\infty}_{l=0}$ such that
$\lim_{l\rightarrow\infty}\sum^n_{i=1}|x_{i}(k_l)|_{X_i}=0$.
Since the consensus is achieved by what we have proven in the first step (\textbf{i}), we have
$$
\lim_{l\rightarrow\infty}\sum^n_{i=1}|x_{i}(k_l)|_{X_j}=0\;
\mbox{for}\; \mbox{all}\; j\in\mathcal{V},
$$
which implies
$\vartheta=\lim_{l\rightarrow\infty}\max_{1\leq i\leq
n}|x_i(k_l)|_{X_0}=0$.
\end{itemize}

This completes the proof. \hfill$\square$

\begin{rem}
If the adjacency matrices are double-stochastic, $\sum^{\infty}_{k=0}\alpha^-_k=\infty$
and $\sum^{\infty}_{k=0}\alpha^+_k\theta_k<\infty$, then the convergence analysis for the optimal consensus can also be performed
by the following procedures.

By summing the two sides in (\ref{177}) over $i=1,...,n$, we obtain
\begin{align}\label{rem}
\sum^n_{i=1}|x_i(k+1)|_{X_0}&\leq (1+\alpha^+_k\tan\theta_k)\sum^n_{i=1}|x_i(k)|_{X_0}
-\sum^n_{i=1}\alpha_{i,k}\Big(|x_i(k)|_{X_0}-\sqrt{|x_i(k)|^2_{X_0}-|x_i(k)|^2_{X_i}}\Big).
\end{align}
Summing the two sides in (\ref{rem}) over $k\geq 0$ and rearranging the terms, we have
\begin{align}\label{rem2}
\sum^{\infty}_{k=0}\alpha^-_{k}\sum^n_{i=1}\Big(|x_i(k)|_{X_0}-&
\sqrt{|x_i(k)|^2_{X_0}-|x_i(k)|^2_{X_i}}\Big)\nonumber\\
&\leq\sum^{\infty}_{k=0}\sum^n_{i=1}\alpha_{i,k}\Big(|x_i(k)|_{X_0}-
\sqrt{|x_i(k)|^2_{X_0}-|x_i(k)|^2_{X_i}}\Big)
\nonumber\\
&\leq\sum^n_{i=1}|x_i(0)|_{X_0}+\sum^{\infty}_{k=0}\alpha^+_k\tan\theta_k\sum^n_{i=1}|x_i(k)|_{X_0}<\infty.
\end{align}

The assumption $\sum^{\infty}_{k=0}\alpha^-_k=\infty$ and (\ref{rem2}) imply that
$$
\liminf_{k\rightarrow\infty}\sum^n_{i=1}\Big(|x_i(k)|_{X_0}-\sqrt{|x_i(k)|^2_{X_0}-|x_i(k)|^2_{X_i}}\Big)=0,
$$
and then
\begin{align}\label{rem4}
\liminf_{k\rightarrow\infty}\sum^n_{i=1}|x_i(k)|_{X_i}=0.
\end{align}

From (\ref{rem2}) we also have
\begin{align}\label{rem3}
\lim_{k\rightarrow\infty}\sum^n_{i=1}\alpha_{i,k}\Big(|x_i(k)|_{X_0}-\sqrt{|x_i(k)|^2_{X_0}-|x_i(k)|^2_{X_i}}\Big)=0
\end{align}
and then for all $i\in \mathcal{V}$,
\begin{align}\label{rem5}
\limsup_{k\rightarrow\infty}\alpha_{i,k}|x_{i}(k)|_{X_i}=0
\end{align}
since if there is $i_0\in \mathcal{V}$ and $\{k_l\}^{\infty}_{l=1}$ such that
$\alpha_{i_0, k_l}|x_{i_0}(k_l)|_{X_{i_0}}\geq \varepsilon$ for all $l$ and some $\varepsilon>0$,
then for all $l$,
\begin{align}
\alpha_{i_0, k_l}\Big(|x_{i_0}(k_l)|_{X_0}-\sqrt{|x_{i_0}(k_l)|^2_{X_0}-|x_i(k_l)|^2_{X_{i_0}}}\Big)
&\geq\alpha_{i_0, k_l}|x_{i_0}(k_l)|_{X_0}-\sqrt{{\alpha}^2_{i_0, k_l}|x_{i_0}(k_l)|^2_{X_0}-\varepsilon^2}>0,\nonumber
\end{align}
which contradicts with (\ref{rem3}). As what have been proved in Theorem \ref{thm1}, (\ref{rem4}) and (\ref{rem5}) imply that
 the optimal consensus is achieved.

\end{rem}

\subsubsection{Proof of Theorem \ref{thm2}}

It is easy to find that if $\theta_k\equiv0$,
the intersection set in (\ref{ang}) is the line segment from $x_i(k)$ to $P_{X_i}(x_i(k))$
and then $P^{sa}_i(k)=P_{X_i}(x_i(k))$. Then the evolution  of the approximate projected consensus
algorithm becomes
\begin{align}\label{25}
x_i(k+1)=\sum_{j\in\mathcal{N}_i(k)}a_{ij}(k)\Big((1-\alpha_{j,k})x_j(k)+\alpha_{j,k}P_{X_j}(x_j(k))\Big),\; i=1,...,n.
\end{align}

We complete the proof by the following two parts.
\begin{itemize}
\item [(i)]
We first prove that if $\sum^{\infty}_{k=0}\alpha^+_k<\infty$,
then there exist initial conditions from which all agents will not converge to $X_0$.
Let $z^*\in R^m$, which shall be selected later, and $x_i(0)=z^*,i=1,...,n$.

By (\ref{25}), $x_i(1)$
can be rewritten as
\begin{align}
x_i(1)&=\sum_{j\in\mathcal{N}_i(0)}a_{ij}(0)\Big((1-\alpha_{j,0}) x_j(0) + \alpha_{j,0} P_{X_j}(x_j(0))\Big)\nonumber\\
&=\sum_{j\in\mathcal{N}_i(0)}a_{ij}(0)(1-\alpha_{j,0})z^*+\sum_{j\in\mathcal{N}_i(0)}a_{ij}(0)\alpha_{j,0}P_{X_0}(z^*)+\Delta_{i0},\nonumber\\
&=(1-\beta_{i,0})z^*+\beta_{i,0}P_{X_0}(z^*)+\Delta_{i0},\nonumber
\end{align}
where $1-\beta_{i,0}=\sum_{j\in\mathcal{N}_i(0)}a_{ij}(0)(1-\alpha_{j,0})$ and
$\Delta_{i0}=\sum_{j\in\mathcal{N}_i(0)}a_{ij}(0)\alpha_{j,0}(P_{X_j}(z^*)-P_{X_0}(z^*))$ with
$|\Delta_{i0}|\leq\alpha^+_0 d^*$ for all $i$.

We also have
\begin{align}
x_i(2)&=\sum_{j\in\mathcal{N}_i(1)}a_{ij}(1)\Big((1-\alpha_{j,1}) x_j(1) + \alpha_{j,1} P_{X_j}(x_j(1))\Big)\nonumber\\
&=\sum_{j\in\mathcal{N}_i(1)}a_{ij}(1)(1-\alpha_{j,1})\Big((1-\beta_{j,0})z^*+\beta_{j,0}P_{X_0}(z^*)\Big)+\Delta_{i1}\nonumber\\
&\quad\quad\quad\quad+\sum_{j\in\mathcal{N}_i(1)}a_{ij}(1)\alpha_{j,1}P_{X_0}\Big((1-\beta_{j,0})z^*+\beta_{j,0}P_{X_0}(z^*)\Big)\nonumber\\
&=(1-\beta_{i,1})z^*+\beta_{i,1}P_{X_0}(z^*)+\Delta_{i1},\nonumber
\end{align}
where $1-\beta_{i,1}=\sum_{j\in\mathcal{N}_i(1)}a_{ij}(1)(1-\alpha_{j,1})(1-\beta_{j,0})$,
the third equality follows from Lemma \ref{lem2} (iii) and $\Delta_{i1}=\Delta^1_{i1}+\Delta^2_{i1}+\Delta^3_{i1}$ with
\begin{align}\label{28}
&\Delta^1_{i1}=\sum_{j\in\mathcal{N}_i(1)}a_{ij}(1)(1-\alpha_{j,1})\Delta_{j0};\nonumber\\
&\Delta^2_{i1}=\sum_{j\in\mathcal{N}_i(1)}a_{ij}(1)\alpha_{j,1}\Big(P_{X_j}(x_j(1))-P_{X_0}(x_j(1))\Big);\nonumber\\
&\Delta^3_{i1}=\sum_{j\in\mathcal{N}_i(1)}a_{ij}(1)\alpha_{j,1}\Big(P_{X_0}(x_j(1))-P_{X_0}\big((1-\beta_{j,0})z^*+\beta_{j,0}P_{X_0}(z^*)\big)\Big).
\end{align}
Now we give an estimate of the upper bound of $\Delta_{i1}$. By Lemma \ref{lem2} (i),
\begin{align}
\big|P_{X_0}(x_j(1))-P_{X_0}\big((1-\beta_{j,0})z^*+\beta_{j,0}P_{X_0}(z^*)\big)\big|&\leq\big|x_j(1)-\big((1-\beta_{j,0})z^*+\beta_{j,0}P_{X_0}(z^*)\big)\big|\nonumber\\
&=|\Delta_{j0}|,\nonumber
\end{align}
which implies that $|\Delta^1_{i1}|+|\Delta^3_{i1}|\leq\max_{1\leq i\leq n}|\Delta_{i0}|\leq\alpha^+_0d^*$ and then
$|\Delta_{i1}|\leq|\Delta^1_{i1}|+|\Delta^3_{i1}|+|\Delta^2_{i1}|\leq(\alpha^+_0+\alpha^+_1)d^*$ for all $i$.

Similarly, we can show by induction that for all $i$ and $k$, $x_i(k+1)$ can be expressed as
\begin{align}\label{29}
x_i(k+1)=(1-\beta_{i,k}) z^*+\beta_{i,k}P_{X_0}(z^*)+\Delta_{ik},
\end{align}
where $|\Delta_{ik}|\leq\sum^k_{l=0}\alpha^+_ld^*$ and $\{\beta_{i,k},i\in \mathcal{V}\}^{\infty}_{k=0}$
satisfies
\begin{align}\label{26}
1-\beta_{i,k}=\sum_{j\in\mathcal{N}_i(k)}a_{ij}(k)(1-\alpha_{j,k})(1-\beta_{j,k-1}).
\end{align}
Based on (\ref{26}), we can show by induction that
\begin{align}\label{27}
1-\beta_{i,k}\geq\prod^k_{l=0}(1-\alpha^+_l)\;\;\mbox{for all}\; i\; \mbox{and}\; k.
\end{align}

It follows from (\ref{29}) and Lemma \ref{lem2} (ii) that
\begin{align}\label{30}
|x_i(k+1)|_{X_0}&\geq\big|(1-\beta_{i,k}) z^*+\beta_{i,k}P_{X_0}(z^*)\big|_{X_0}-|\Delta_{ik}|.
\end{align}
Moreover, for a convex set $K$ and any $0\leq\lambda\leq1$,
\begin{align}\label{31}
\big|(1-\lambda) x+ \lambda P_K(x)\big|_{K}
&=\big|(1-\lambda) x+\lambda P_K(x)- P_K\big((1-\lambda) x+ \lambda P_K(x)\big)\big|\nonumber\\
&=\big|(1-\lambda) x+\lambda P_K(x)- P_K(x)\big|\nonumber\\
&=(1-\lambda)|x|_{K},
\end{align}
where the second equality follows from Lemma \ref{lem2} (iii).
By taking $\lambda=\beta_{i,k}$ in (\ref{31}),
we have $$\big|(1-\beta_{i,k}) z^*+\beta_{i,k}P_{X_0}(z^*)\big|_{X_0}=(1-\beta_{i,k})|z^*|_{X_0}.$$
Combining the last equality, (\ref{30}) and (\ref{27}), we obtain
\begin{align}\label{32}
|x_i(k+1)|_{X_0}&\geq\prod^k_{l=0}(1-\alpha^+_l)|z^*|_{X_0}-\sum^k_{l=0}\alpha^+_ld^*.
\end{align}

Taking the inferior limit on the two sides in (\ref{32}), we have for all $i$,
$$
\liminf_{k\rightarrow\infty}|x_i(k)|_{X_0}\geq\prod^\infty_{l=0}(1-\alpha^+_l)|z^*|_{X_0}-\sum^\infty_{l=0}\alpha^+_ld^*,
$$
which is positive provided that
\begin{align}\label{33}
|z^*|_{X_0}>\frac{\sum^\infty_{l=0}\alpha^+_ld^*}{\prod^\infty_{l=0}(1-\alpha^+_l)},
\end{align}
where $\prod^\infty_{l=0}(1-\alpha^+_l)>0$ since $0\leq\alpha^+_l<1$ for all $l$ and $\sum^\infty_{l=0}\alpha^+_l<\infty$.
Thus, all agents can not achieve an optimal consensus for all initial conditions
 satisfying (\ref{33}).

\item [(ii)]
In fact, if $\sum^{\infty}_{k=0}\alpha^+_k<\infty$, there is $y^*=y^*(z^*)\not\in X_0$
such that $\lim_{k\rightarrow\infty}x_i(k)=y^*$ for all $i$ provided $z^*$ satisfies (\ref{33}).

Denote $\bar d=\sup_{1\leq i\leq n,k\geq 0}|x_i(k)|_{X_i}$. By (\ref{25}), we have  for all $i$ and $k$,
\begin{align}\label{41}
x_i(k+1)=\sum_{j\in\mathcal{N}_i(k)}a_{ij}(k)x_j(k)+\Gamma_{i,k},
\end{align}
 where
$\Gamma_{i,k}=\sum_{j\in\mathcal{N}_i(k)}a_{ij}(k)\alpha_{jk}\big(P_{X_j}(x_j(k))-x_j(k)\big)$
with $|\Gamma_{i,k}|\leq \bar d\alpha^+_k$.

Take $i_0\in\mathcal{V}$. It follows from (\ref{41}) that for all $j$ and $k$,
\begin{align}\label{42}
|x_j(k+1)-x_{i_0}(k)|\leq\max_{1\leq p,q\leq n}|x_p(k)-x_q(k)|+\max_{1\leq r\leq n}|\Gamma_{r,k}|.
\end{align}
Then
\begin{align}
&\ \ \ \ |x_{i_0}(k+2)-x_{i_0}(k)|\nonumber\\
&=\Bigg|\sum_{j\in\mathcal{N}_{i_0}(k+1)}a_{i_0j}(k+1)x_j(k+1)+
\sum_{j\in\mathcal{N}_{i_0}(k+1)}a_{i_0j}(k+1)\Gamma_{j,k+1}-x_{i_0}(k)\Bigg|\nonumber\\
&\leq\max_{1\leq j\leq n}|x_j(k+1)-x_{i_0}(k)|+\max_{1\leq r\leq n}|\Gamma_{r,k+1}|\nonumber\\
&\leq\max_{1\leq p,q\leq n}|x_p(k)-x_q(k)|+\max_{1\leq r\leq n}|\Gamma_{r,k}|+\max_{1\leq r\leq n}|\Gamma_{r,k+1}|\nonumber\\
&\leq \max_{1\leq p,q\leq n}|x_p(k)-x_q(k)|+\bar d(\alpha^+_k+\alpha^+_{k+1}).\nonumber
\end{align}
We can similarly show by induction that for all $k$ and $l$,
\begin{align}|x_{i_0}(k+l)-x_{i_0}(k)|&\leq \max_{1\leq p,q\leq n}|x_p(k)-x_q(k)|+\bar d\sum^{k+l-1}_{s=k}\alpha^+_s\nonumber\\
&\leq \max_{1\leq p,q\leq n}|x_p(k)-x_q(k)|+\bar d\sum^{\infty}_{s=k}\alpha^+_s.
\end{align}

Since $\lim_{k\rightarrow\infty}\alpha^+_k=0$ and
$\{x_i(k)\}^{\infty}_{k=0}$ is bounded for all $i$, the consensus is achieved by Lemma \ref{lem5}.
Combining the consensus and the boundedness of $\{x_i(k)\}^{\infty}_{k=0}$, there exist a subsequence $\{k_l\}^{\infty}_{l=0}$ and
$y^*$ such that $\lim_{l\rightarrow\infty}x_i(k_l)=y^*$ for all $i$. Therefore, since
$\sum^{\infty}_{k=0}\alpha^+_k<\infty$, for any $\varepsilon>0$, there exists $l_0=l_0(\varepsilon)$ such that $|x_i(k_l)-x_j(k_l)|\leq \varepsilon/2$ and $\bar d\sum^{\infty}_{s=k_l}\alpha^+_s\leq\varepsilon/2$ for all $i,j$ and $l\geq l_0$. Thus,
for $l\geq l_0$ and $k\geq k_l$,
$$|x_{i_0}(k)-x_{i_0}(k_l)|\leq\max_{1\leq p,q\leq n}|x_p(k_l)-x_q(k_l)|+\bar d\sum^{\infty}_{s=k_l}\alpha^+_s\leq\varepsilon,$$
which implies that $\lim_{l\rightarrow\infty}x_{i_0}(k)=y^*$ since $\varepsilon$ can be arbitrarily small. By what we have proven in the first part (i), $y^*\not\in X_0$. The desired conclusion follows from the fact that $i_0$ is taken  arbitrarily.
\end{itemize}

The proof is completed. \hfill$\square$
\section{Critical Angle Error}
The boundedness of system states  plays a key role for various optimization methods \cite{nedic3,nedic1}.
In this section, we consider the effect of the angle error $\theta_k$ on the boundedness of the states $\{x_i(k),i\in \mathcal{V}\}^{\infty}_{k=0}$ generated by the approximate projected consensus
algorithm.

Suppose $\alpha_{i,k}\equiv1$ and $\theta_k\equiv\theta$ with $0<\theta<\pi/2$.
First the following conclusion  shows that when $\theta<\pi/4$, the trajectories of the algorithm are uniformly bounded with respect to all initial conditions.
 \begin{pro}
\label{pro1}
Suppose {\bf A1}, {\bf A2}, {\bf A5} hold
and $0<\theta<\pi/4$. Then we have
$$
\sup_{x(0)}\limsup_{k\rightarrow\infty}|x_i(k)|_{X_0}<\infty
$$
for $i=1,\dots,n$.
 \end{pro}
 Proof:
Recall $d^*=\sup_{\omega_1,\omega_2\in\bigcup^n_{i=1}X_i}|\omega_1-\omega_2|$.
We claim that for all $i$ and all initial conditions $x(0)$,
\begin{align}
\limsup_{k\rightarrow\infty}|x_i(k)|_{X_0}\leq \frac{2d^*}{1-\tan\theta},\nonumber
\end{align}
which implies the conclusion.

Based on (\ref{435}), (\ref{pro}) and the definition of $d^*$, we always have
\begin{align}\label{34}
|P^{sa}_i(k)|_{X_0}&\leq\tan\theta|x_i(k)|_{X_i}+|P_{X_i}(x_i(k))|_{X_0}\nonumber\\
&\leq\tan\theta|x_i(k)|_{X_0}+d^*,
\end{align}
Furthermore, in the case of $|x_i(k)|_{X_0}\geq 2d^*/(1-\tan\theta)$, by (\ref{34}) we further obtain
\begin{align}\label{35}
|P^{sa}_i(k)|_{X_0}&\leq \tan\theta|x_i(k)|_{X_0}+d^*\nonumber\\
&\leq\Big(1-\frac{1-\tan\theta}{2}\Big)|x_i(k)|_{X_0}.
\end{align}

We next consider $h(k):=\max_{1\leq i\leq n}|x_i(k)|_{X_0}$ by the following two cases:
\begin{itemize}
\item [(i)] $h(k)\leq 2d^*/(1-\tan\theta)$.
By Lemma \ref{lem1} and (\ref{34}), we have for all $i$,
\begin{align}\label{36}
|x_i(k+1)|_{X_0}&\leq\sum_{j\in\mathcal{N}_i(k)}a_{ij}(k)|P^{sa}_j(k)|_{X_0}\nonumber\\
&=\sum_{j\in\mathcal{N}_i(k)}a_{ij}(k)\big(\tan\theta|x_j(k)|_{X_0}+d^*\big)\nonumber\\
&\leq\frac{2d^*\tan\theta}{1-\tan\theta}+d^*\nonumber\\
&\leq \frac{2d^*}{1-\tan\theta}.
\end{align}
Therefore, $h(k+1)\leq2d^*/(1-\tan\theta)$.

\item [(ii)] $h(k)> 2d^*/(1-\tan\theta)$.
Define $\mathcal{V}_1(k)=\{i|\;|x_j(k)|_{X_0}\leq 2d^*/(1-\tan\theta)$ for all $j\in\mathcal{N}_i(k)\}$
and $\mathcal{V}_2(k)=\mathcal{V}\backslash\mathcal{V}_1(k)$, where $\mathcal{V}_2(k)$ is nonempty.
\begin{itemize}
\item [1)]
If $i\in\mathcal{V}_1(k)$, by similar analysis with (\ref{36}), we obtain $|x_i(k+1)|_{X_0}\leq 2d^*/(1-\tan\theta)$.
\item [2)]
If $i\in\mathcal{V}_2(k)$, we can define $\mathcal{N}^1_i(k)=\{j|\;j\in\mathcal{N}_i(k),|x_j(k)|_{X_0}\leq 2d^*/(1-\tan\theta)\}$
and $\mathcal{N}^2_i(k)=\mathcal{N}_i(k)\backslash\mathcal{N}^1_i(k)$, where $\mathcal{N}^2_i(k)$ is nonempty.

Again by Lemma \ref{lem1}, we have
\begin{align}\label{37}
|x_i(k+1)|_{X_0}&\leq\sum_{j\in\mathcal{N}^1_i(k)}a_{ij}(k)|P^{sa}_j(k)|_{X_0}
+\sum_{j\in\mathcal{N}^2_i(k)}a_{ij}(k)|P^{sa}_j(k)|_{X_0}.
\end{align}

For $j\in\mathcal{N}^1_i(k)$, by (\ref{34}) we have
$|P^{sa}_j(k)|_{X_0}\leq\tan\theta|x_j(k)|_{X_0}+d^* \leq2d^*\tan\theta/(1-\tan\theta)+d^*\leq2d^*/(1-\tan\theta)$.
As a result, by (\ref{35}) and (\ref{37}) we have
\begin{align}\label{38}
|x_i(k+1)|_{X_0}
&\leq \frac{2d^*}{1-\tan\theta}\sum_{j\in\mathcal{N}^1_i(k)}a_{ij}(k)
+\Big(1-\frac{1-\tan\theta}{2}\Big)\sum_{j\in\mathcal{N}^2_i(k)}a_{ij}(k)|x_j(k)|_{X_0}.
\end{align}
Since $|x_j(k)|_{X_0}\geq 2d^*/(1-\tan\theta)$ for all $j\in\mathcal{N}^2_i(k)$,
$\max_{j\in\mathcal{N}_i(k)}|x_j(k)|_{X_0}\geq 2d^*/(1-\tan\theta)$. Thus, it follows from (\ref{38}) that
\begin{align}
|x_i(k+1)|_{X_0}
&\leq\Big(\sum_{j\in\mathcal{N}^1_i(k)}a_{ij}(k)+\big(1-\frac{1-\tan\theta}{2}\big)\sum_{j\in\mathcal{N}^2_i(k)}a_{ij}(k)\Big)
\max_{j\in\mathcal{N}_i(k)}|x_i(k)|_{X_0}\nonumber\\
&\leq\Big((1-\eta)+\eta\big(1-\frac{1-\tan\theta}{2}\big)\Big)h(k+1)\nonumber\\
&=\Big(1-\frac{\eta(1-\tan\theta)}{2}\Big)h(k+1),\nonumber
\end{align}
where $0<1-\eta(1-\tan\theta)/2<1$ and $\eta$ is the lower bound of weights in {\bf A2}.
\end{itemize}
Therefore, based on the two cases (1) and (2), we show that if $h(k)> 2d^*/(1-\tan\theta)$, then
$$h(k+1)\leq\max\big\{2d^*/(1-\tan\theta), \big(1-\eta(1-\tan\theta)/2\big) h(k)\big\}.$$
\end{itemize}
Combining the two cases (i) and (ii), we have
$$h(k+1)\leq\left\{
\begin{array}{ll}
2d^*/(1-\tan\theta),\qquad\qquad\qquad\qquad\qquad\qquad\qquad\;\;\mbox{if}\;h(k)\leq2d^*/(1-\tan\theta);\\
\max\big\{2d^*/(1-\tan\theta), \big(1-\eta(1-\tan\theta)/2\big) h(k)\big\},\;\mbox{otherwise}
\\
\end{array}
\right.
$$
Thus, the conclusion follows.
\hfill$\square$

We next investigate the non-conservativeness of $\pi/4$ in Proposition \ref{pro1}. We just focus on a special case with only one
 node in the network. Its set is denoted as $X_\ast$. We denote the states of the node as $\{x_\ast(k)\}^{\infty}_{k=0}$ driven by the approximate projected consensus
algorithm:
$$
x_*(k+1)\in \mathscr{P}^{sa}_{X_*}(x_*(k), \theta),
$$
where $\mathscr{P}^{sa}_{X_*}(x_*(k), \theta)=\mathbf{C}_{X_*}(x_*(k),\theta)\bigcap \mathbf{H}_{X_*}(x_*(k))$.

The following conclusion holds.
 \begin{pro}
\label{pro2}
Suppose $\theta=\pi/4$.

(i) For any compact convex set $X_*$ and any initial condition $x_*(0)\in R^m$, we have
$$\limsup_{k\rightarrow\infty}|x_*(k)|_{X_*}\leq |x_*(0)|_{X_*};$$

(ii) There exists an approximate projection sequence $\{P^{sa}_*(k)\}^{\infty}_{k=0}$,  $P^{sa}_*(k)\in\mathscr{P}^{sa}_{X_*}(x_*(k), \pi/4)$ such that

\quad (ii.1)\; $\limsup_{k\rightarrow\infty}|x_*(k)|_{X_*}=0$ when $X_*$ is a ball with radius $r>0$;

\quad (ii.2)\; $\limsup_{k\rightarrow\infty}|x_*(k)|_{X_*}= |x_*(0)|_{X_*}$ when $X_*$ is a single point.
\end{pro}
Proof:
(i)
The conclusion follows from that
$$
|x_*(k+1)|_{X_*}\leq\big|x_*(k+1)-P_{X_*}(x_*(k))\big|\leq \tan\theta|x_*(k)|_{X_*}=|x_*(k)|_{X_*}.
$$

(ii) We just select $\{P^{sa}_*(k)\}^{\infty}_{k=0}$ for which ${\rm Ang}\big(P^{sa}_*(k)-x_*(k),P_{X_*}(x_*(k))-x_*(k)\big)=\pi/4$ for all $k$
satisfying the definition (\ref{ang}) such that the solution
of the algorithm with $\alpha_{i,k}\equiv1$ and $\theta_k\equiv\pi/4$ satisfies the following two cases.

(ii.1) It is easy to find that for all $k$,
\begin{align}\label{39}
|x_*(k+1)|_{X_*}=\sqrt{|x_*(k)|^2_{X_*}+r^2}-r,
\end{align}
which implies that the sequence $\{|x_*(k)|_{X_*}\}^{\infty}_{k=0}$ is non-increasing and then converges to some nonnegative
number $\mu$. Letting $k\rightarrow\infty$ in (\ref{39}), we have $\mu=\sqrt{\mu^2+r^2}-r$ and then $\mu=0$.

\begin{figure}[!htbp]
\centering
\includegraphics[width=3.0in]{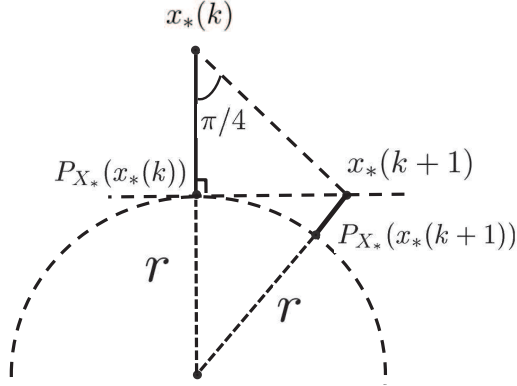}
\caption{Approximate projection with respect to a ball.}
\end{figure}

(ii.2) The conclusion is straightforward.
\hfill$\square$

We present another result for the case when $\theta>\pi/4$ to reveal that, in this case, the node states will be unbounded as long as the
distance between the initial condition and $X_*$ is larger than a certain threshold.
 \begin{pro}
\label{pro3}
Suppose $\theta>\pi/4$. Then for
any compact convex set $X_*$, there exists an approximate projection sequence $\{P^{sa}_*(k)\}^{\infty}_{k=0}$ such that
$$\limsup_{k\rightarrow\infty}|x_*(k)|_{X_*}=\infty$$
for all initial conditions satisfying $|x_*(0)|_{X_*}>\sup_{\omega_1,\omega_2\in X_*}|\omega_1-\omega_2|/(\tan\theta-1)$.
\end{pro}
Proof:
Select $\{P^{sa}_*(k)\}^{\infty}_{k=0}$ for which ${\rm Ang}\big(P^{sa}_*(k)-x_*(k),P_{X_*}(x_*(k))-x_*(k)\big)=\theta$ for all $k$.
For all $k$ we have
\begin{align}\label{40}
|x_*(k)|_{X_*}&\geq\big|x_*(k)-P_{X_*}(x_*(k-1))\big|-\big|P_{X_*}(x_*(k-1))-P_{X_*}(x_*(k))\big|\nonumber\\
&=\tan\theta|x_*(k-1)|_{X_*}-\big|P_{X_*}(x_*(k-1))-P_{X_*}(x_*(k))\big|\nonumber\\
&\geq\tan\theta|x_*(k-1)|_{X_*}-\sup_{\omega_1, \omega_2\in X_*}|\omega_1-\omega_2|.
\end{align}

At this point, we need the following conclusion:
consider a nonnegative sequence $\{z_k\}^{\infty}_{k=0}$ with $z_{k+1}\geq (\tan\theta)z_k -\hat d$ and $\theta>\pi/4$.
Then $\lim_{k\rightarrow\infty}z_k=\infty$ if $(\tan\theta-1)z_0-\hat d>0$.
Note that $z_{1}-z_0\geq(\tan\theta-1)z_0-\hat d>0$ and then
$z_2-z_1\geq(\tan\theta-1)z_1-\hat d\geq (\tan\theta-1)z_0-\hat d>0$. We can show similarly by induction that
$z_{k+1}-z_k\geq(\tan\theta-1)z_0-\hat d$ for all $k$.
The conclusion follows from (\ref{40}) and the above conclusion.
\hfill$\square$

Combining Propositions \ref{pro1}, \ref{pro2} and \ref{pro3}, we see that $\pi/4$ is a critical value of the angle error
in the approximate projection regarding maintaining bounded iterative states.
If $\theta<\pi/4$, the system trajectories are uniformly bounded; if $\theta>\pi/4$, the trajectories diverge for
a special case with one single node and particular approximate projection points; if $\theta=\pi/4$, the trajectories of the algorithm with one node are bounded
(no longer uniformly with respect to initial conditions) and the property of the trajectories highly depend on the shape of the convex set.

\section{Numerical Example}
In this section, we provide a numerical example which shows that
the approximate projected consensus
algorithm presented in this paper may lead to a faster convergence
than the projected consensus algorithm presented in \cite{nedic2}.

Consider a network with three nodes $1$,
$2$ and $3$. Suppose $x_i(k)\in R^2, i=1,2,3$. Let $X_1$, $X_2$ and $X_3$
be three unit disks with centers $(1, 0), (-1, 0)$ and $(0,-1)$, respectively.
Set $X_i$ corresponds to node $i$ for $i=1,2,3$. The communication graph is completely connected.
Take $a_{11}=a_{22}=a_{33}=0.5$ and all other weights are 0.25.
Here $\theta_k\equiv0$, which is a special case of the approximate projected consensus algorithm.

The three agents have the same initial condition $x_1(0)=x_2(0)=x_3(0)$.
We consider 2500 initial conditions
$\big\{(-1.96+0.08p,-1.96+0.08q)|\;0\leq p\leq 49,0\leq q\leq 49\big\}$
equally spaced over the square $\{(z_1,z_2)|\;|z_1|\leq2,|z_2|\leq2\}$.
The initial conditions from which the approximate projected consensus
algorithm with $\alpha_{i,k}\equiv0.5$ converges faster
than the projected consensus algorithm presented in \cite{nedic2} (compare their distance function $h(k)=\max_{1\leq i\leq 3}|x_i(k)|_{X_0}$ at time 2000) are labeled with stars in Figure 4. The decay process of $h(k)$ of a concrete example with initial condition $x_1(0)=x_2(0)=x_3(0)=(1.8, 0.8)$ is given in Figure 5 (drawn from time 20), from which
we can see that the approximate projected consensus
algorithm has better
performance from the viewpoint of convergence speed.

\begin{figure}[!htbp]
\centering
\includegraphics[width=4.0in]{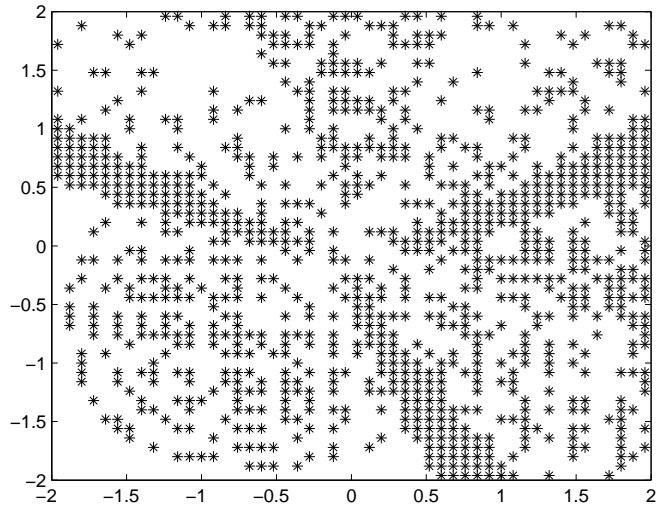}
\caption{Approximate projected consensus
algorithm converges faster than projected consensus
algorithm for the initial conditions labeled stars.}
\end{figure}

\begin{figure}[!htbp]
\centering
\includegraphics[width=4.0in]{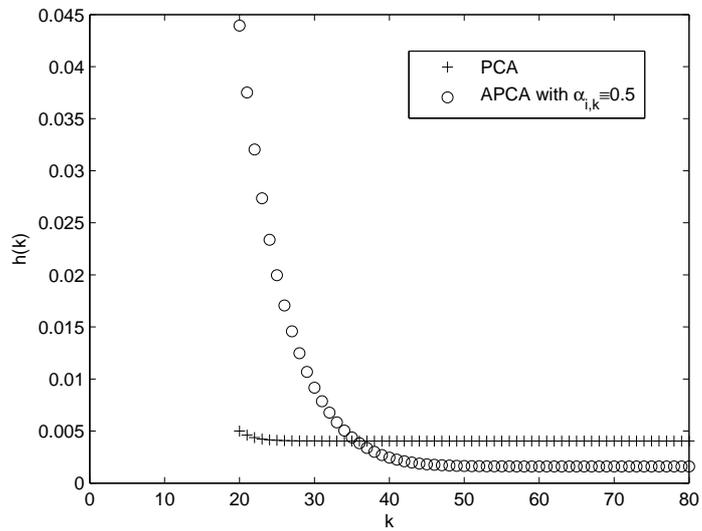}
\caption{A typical solution for which approximate projected consensus
algorithm (APCA) converges faster than projected consensus
algorithm (PCA).}
\end{figure}

\section{Conclusions}

In this paper, we presented an approximate projected consensus algorithm  for a network to cooperatively compute the intersection of a serial of  convex sets, each of which is known only to a particular node.  We allowed  each node to only compute an approximate projection which locates in the intersection of the convex cone generated by the current state and all directions with the exact projection direction less than some angle and the half-space containing the current state with its boundary is a supporting hyperplane to its own set at its exact projection point onto its set. Sufficient and/or necessary conditions were obtained for the considered algorithm on how much projection accuracy is required to ensure a global consensus within the intersection set, under the assumption that the communication graph is uniformly jointly strongly connected. We also showed  that $\pi/4$ is a threshold for the angle error in the projection approximation to ensure a bounded solution for iterative projections. A numerical example was also given indicating that the approximate projected consensus
algorithm sometimes achieves better performance than the exact projected consensus algorithm. This implied that, individual optimum seeking may not be so important for optimizing the collective objective.

\end{document}